\documentclass[useAMS,usenatbib,referee]{mn2e}
\usepackage[cp866]{inputenc}
\usepackage[english]{babel}
\usepackage{graphicx}
\usepackage{amsmath}
\usepackage{epsf}

\def\la{\;\raise0.3ex\hbox{$<$\kern-0.75em\raise-1.1ex\hbox{$\sim$}}\;}
\def\ga{\;\raise0.3ex\hbox{$>$\kern-0.75em\raise-1.1ex\hbox{$\sim$}}\;}

\title[The redshift distribution]
{The redshift distribution \\
of absorption-line systems in QSO spectra}

\author[A.~I.~Ryabinkov, A.~D.~Kaminker, D.~A.~Varshalovich]
{A.~I.~Ryabinkov,
A.~D.~Kaminker \thanks{Send offprint request to: A.~D.~Kaminker},
and D.~A.~Varshalovich  \\ 	 
Ioffe Physico-Technical Institute,
Politekhnicheskaya 26, 194021 St.~Petersburg, Russia,  \\
e-mail: calisto@rbcmail.ru, [kam, varsh]@astro.ioffe.ru}

\begin{document}

\date{Accepted by MNRAS 2006 January 30. Received 2006 December 5; 
in original form 2006  April 6}

\pagerange{\pageref{firstpage}--\pageref{lastpage}} \pubyear{2007}

\maketitle

\label{firstpage}

\begin{abstract}
A statistical analysis of the
space-time distribution of
absorption-line systems (ALSs) observed in QSO spectra
within the cosmological redshift interval $z=0.0$--3.7
is carried out on the base of our 
catalog of absorption systems \citep{rkv03}.
We confirm our previous conclusion 
that the $z$-distribution of absorbing matter
contains non-uniform component displaying
a pattern of statistically significant
alternating maxima (peaks)
and minima (dips). 
Using the wavelet transformation
we determine the positions of the maxima and minima
and estimate their statistical significance.
The positions of the maxima and minima
of the $z$-distributions obtained for different
celestial hemispheres turn out to be
weakly sensitive to orientations of the hemispheres.
The data reveal a regularity (quasi-periodicity)
of the sequence of the peaks and dips 
with respect to 
some rescaling functions of $z$.
The same periodicity was found  
for the one-dimensional
correlation function calculated for the sample of
the ALSs
under investigation.
We assume the existence of a regular
structure in the distribution of absorption matter,
which is not only spatial but 
also
temporal
in nature with characteristic time 
varying within the interval
150--650~Myr
for the cosmological model applied.
\end{abstract}


%
\section{Introduction}
\label{intro}
We continue our previous studies
(e.g.,  \citealt*{krv00}, hereafter Paper~I)
of the space-time distribution
of absorption-line systems (ALSs)
imprinted in spectra of quasars (QSOs).
Actually, ALSs
contain basic information on
the distribution of matter
between the observer and  QSOs
as well as 
on physical processes occurred 
in different epochs
of the cosmological evolution.
In the first stage 
(e.g., Paper~I) 
we used as indicators of matter
only  data on
the resonance absorption
doublets of C~IV and Mg~II
observed in QSO spectra
at cosmological redshifts $z = 0.2-3.2$
paying special
attention to
diminishing 
of possible selection effects.
Then, we explored considerably larger
number of ALSs 
(\citealt*{rkv01}; hereafter Paper~II)
basing on the catalog by
\citet*{jhb91}
within 
the extended range $z = 0.0-3.7$.
It was shown 
that the $z$-distribution of ALSs displays
a pattern of alternating maxima (peaks)
and minima (dips) which
are statistically significant
against a smooth dependence (trend).
It is essential that the positions of
the peaks and dips turned out to be
independent (within statistical uncertainties)
of observation directions.
This suggested that the derived distribution
of absorption matter is not only spatial
but 
also 
temporal in nature.

We present
the results of an extended statistical analysis
performed  by different methods
on the base of our
catalog of absorption systems
\citep{rkv03}.
\begin{figure*}    
\centering
\includegraphics[width=120mm,  bb=75  485  525  770, clip]{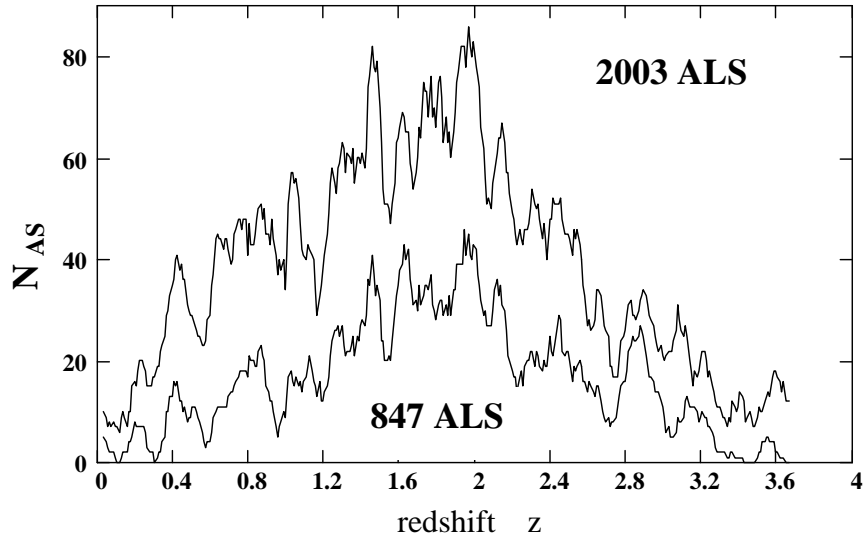}
\vspace{-0.2cm}
\caption{
$z$-distributions
of absorption-line systems observed in QSO spectra 
within the redshift interval $z=0.0$ -- 3.7:
(i) 847 systems are taken      
from the catalogue by   
\citet{jhb91} and 
(ii) 2003 systems --     
from the catalogue by \citet{rkv03} 
(see text).}
\label{AS91-03}   
\end{figure*}
Fig.\ \ref{AS91-03} demonstrates
two $z$-distributions  
$N_{\rm AS}(z)$ of ALSs
containing absorption lines of heavy elements
within the redshift
interval $z=0.0$ -- 3.7. One of them is obtained
using  847 systems from
the catalog by  \citet{jhb91}
and the second one is based on 
2003  systems 
registered in the spectra of 661 QSOs 
(emission redshifts within
$z_e=$0.29 -- 4.9) from the
catalog by \citet{rkv03}.  
All ALSs under consideration 
comprise lines of heavy elements 
and may include up to 20 -- 30 absorption lines 
detected in different regions of QSO spectra
(predominantly within the interval  
3000 -- 8000 \AA). 
We exclude ALSs 
consisting 
only of neutral hydrogen lines 
as well as 
damped Ly$\alpha$ absorption systems (DLA).     
All redshifts $z_j$ 
registered in spectra of each QSO
and fallen into the velocity interval
$\, \delta v =$500 km~s$^{-1}$
are treated as a single absorption system
with an averaged redshift
$z_a=(\sum_{j=1}^{n_a} z_j)/n_a$,
where $n_a$ is a number of redshifts 
within the interval.  
Both distributions in Fig.~\ref{AS91-03}
are obtained using so-called
sliding-average approach  which represents a set of
consecutive displacements of the averaging bin
$\Delta_{\rm z}=0.071$  along $z$-axis
with a step $\delta_{\rm z}=0.01$.

A comparison of the two $z$-distributions
(Fig.\ \ref{AS91-03}) 
reveals 
similar patterns of
the peaks and dips 
relatively smooth curves.
The positions of majority of the peaks and dips
remain the same after the extension of statistics
and some of them become more significant.
There are only a few exceptions 
concerned with 
splitting of some 
initially single
peaks into double ones
and  an appearance of new peaks (see Table\ 1).

In Papers~I and II
the statistical analysis of ALSs was performed
using the averaging bin
$\Delta_{\rm z}=0.071$.
This value was
chosen with employing   
the $\chi^2$-criterion
as a measure of
the most significant deviations
of the $z$-distributions 
obtained for different values 
$\Delta_{\rm z}$ 
from the hypothesis of the uniform  distribution
(with use of the trend subtraction). 
On this way  the results of statistical
considerations  
could be sensitive to  
artificial non-uniformities
induced by a
separation of the
whole sample of the redshifts into bins.
To exclude sensitivity of our results
to the effects of the averaging bins
we employ in this paper 
predominantly out-of-bin statistical technique.
In particular, in Section\ 2
we apply a continuous
wavelet transform to the whole sample of
ALS redshifts to study in details
the peak-and-dips sequence. 

In Section\ 3,
we demonstrate  isotropy  of
the $z$-distribution, i.e.,
approximate independence 
the peak/dip positions 
of observation directions.
In Section\ 4,
we focus on a presumable 
regularity (quasi-periodicity) 
of the $z$-distribution.
In Section\ 5, 
we examine properties of the one-dimensional
correlation function. 
In Section\ 6,  a  polemic 
on the presence of periodicities 
in the distribution of QSO redshifts 
is outlined    
and possible selection effects 
in our analysis 
are briefly discussed. 
Conclusions and 
a short discussion of 
possible interpretations
of the results 
are considered in Section\ 7.
  
\section{Wavelet transform and non-uniformity of the $z$-distribution}
\label{wavelet}
In this section, we use the wavelet transform for 
the analysis of the consecution 
of all redshifts
under consideration 
within the interval $z=0.0$ -- 3.7 
without subdividing 
the $z$-points
into certain 
statistical bins.     
The wavelet transform is a way 
to disclose a difference of 
the redshift distribution 
from a smooth  dependence   
and to reveal a set of 
condensations and depressions
(peaks and dips) beyond 
the Poisson  fluctuations.  
Actually, the wavelet analysis is effective
when a signal has no evident periodicity
and contains different sets of wave number (frequency) 
components in different space (time) regions.
This method appears to be appropriate  
for the study of the non-uniformities
visible in Fig.\  \ref{AS91-03} and  allows us 
to estimate  
statistical  significance of the peaks and dips 
in the $z$-distribution of ALSs.
Detailed information on the wavelet analysis and its
applications in physics and astrophysics  
may be found in numerous books, reviews, and papers
(e.g., \citealt{c92}, \citealt{a96}, 
\citealt*{din01}  
and references therein).
 
We apply so called ``Mexican hat'' wavelet:
%
%
%
\begin{equation}
   \psi_{K_x}(x, y)    =  A  \left[ 1 - u^2 \right]
          \exp \left[ - {u^2 \over 2} \right]; \, \, \, \, \, \,
             u  \equiv u(x, y) = (x-y) K_x,
\label{g(xy)}
\end{equation}
where $x$ is the current one-dimensional coordinate,
$y$ is the {\it translation} of the wavelet along the 
coordinate axis,  $K_x$ is
an analog of the wave number for one-dimensional
Fourier transform, 
and  $A=A(K_x)$ is a normalizing function. 
Usually, the value $1/K_x$
is called a {\it dilation} or a scale parameter.
The wavelet (\ref{g(xy)}) is proportional to the second
derivative of the Gaussian distribution
$\exp [- u^2/2]/\sqrt{2\pi}$  taken with sign $(-)$.
The zero-momentum of Eq.~(\ref{g(xy)}) 
$\int {\rm d}x \, \psi_{K_x}(x, y)=$
$\int {\rm d}y \, \psi_{K_x}(x, y)$
is equal to zero
in accordance with the 
admissibility condition (e.g.,  \citealt{c92}).

Let we have a signal $F(x)$,  where
$F(x) \in {\large L}^2$-space of normalized functions
determined over the real axis $x$.
Then, the coefficients of the  continuous 
wavelet transform are
\begin{equation}
       S(K_x, y) = \int  {\rm d}x \, F(x) \psi_{K_x} (x,y),
\label{S(y)}
\end{equation}
where the integration is carried out over the real axis.
We obtain two-dimensional scanning of the signal,
the wave number $K_x$ and the coordinate $y$ being
independent variables.

In this paper,  we obtain the coefficients 
(\ref{S(y)}) in an unconventional
manner when the signal $F(z)$ is treated as the consecution
of the discrete redshifts $z_i\, $  ($i=1,2,...$)
inside the considered interval (similar approach 
was suggested by \citealt{slb93}). 
We approximately present $F(z)$ as
the sum of normalized delta functions  localizing 
positions $z_i$ of all  points  in the sample:
%
%
\begin{equation}
         F(z) \approx \sum_{i=1}^{{\rm N}_{\rm tot}}
              {2\pi \over {\cal K}_z^{\rm m}}
             \, \delta(z-z_i); \, \, \, \, \, \, \, \, \, \,
            \int_{-\infty}^{\infty} {\rm d}z \,
            |F(z)|^2 \approx {2\pi \over {\cal K}_z^{\rm m}}
            \, {\rm N}_{\rm tot},
\label{F(z)}
\end{equation}
where $ {\cal K}_z^{\rm m} \gg 2 \pi$ is 
an upper boundary 
of all 
wave numbers $(K_z \leq {\cal K}_z^{\rm m})$ involved
in the analysis,
${\rm N}_{\rm tot}$ - is a number
of all absorption systems
in the sample, in our case ${\rm N}_{\rm tot}=2003$. 

Following  Eqs.~(\ref{S(y)}) and  (\ref{F(z)}),
the wavelet transform of the function $F(z)$
yields
the sum over all redshifts:
\begin{equation}
         S(K_z, z)  =  
        {2\pi \over {\cal K}_z^{\rm m}} \,  
	\sum_{i=1}^{{\rm N}_{\rm tot}} \psi_{K_z} (z_i, z) = 
        A(K_z) \, \sum_{i=1}^{{\rm N}_{\rm tot}}
                \left[1-u_i^2 \right]
        \exp \left[ - {u_i^2 \over 2} \right],
\label{S(z)}
\end{equation}
where   $u_i  \equiv u(z_i,\, z) = (z_i-z) K_z$,  
the variables $K_x$ and $y$ in Eq.~(\ref{S(y)}) are replaced by 
$K_z$ and $z$, respectively;  
in the second equality of Eq.~(\ref{S(z)})
the factor of 
$2\pi/{\cal K}_z^{\rm m}$ 
is included in the normalizing function $A(K_z)$.

Under the assumption
of the uniform distribution of points $z_i$
over the interval $z=0.0$ -- 3.7 the value $S(K_z, z)$
represents 
a random function obeying approximately to the Gaussian
distribution
(${\rm N}_{\rm tot} \gg 1$). 
The mean value of the distribution 
$\mu[S(K_z, z)]$ is equal to zero:
\begin{equation}
      \mu[S(K_z, z)] = {1 \over Z_{\rm max}}  
      \sum_{i=1}^{{\rm N}_{\rm tot}} \, \int_0^{Z_{\rm max}}
        {\rm d}z_i \, \psi_{K_z}(z_i,\ z) = 0,
\label{mu0}
\end{equation}
where the upper limit is
$Z_{\rm max}=3.7$
and the mean squared deviation
$\sigma(S(K_z, z))$  may be put equal to unity:
\begin{equation}
      \sigma^2[S(K_z, z)] = {1 \over Z_{\rm max}}  
      \sum_{i=1}^{{\rm N}_{\rm tot}} \, \int_0^{Z_{\rm max}}
        {\rm d}z_i \, \psi_{K_z}^2(z_i,\ z) = 1.
\label{sigma2}
\end{equation}
The latter condition  allows one to determine
$A=A(K_z)$ (in general case $A=A(K_z, z)$). 
Using  Eqs.~(\ref{S(z)}), (\ref{sigma2})
and neglecting terms of the orders of  
$\exp(-K_z^2 z^2) \ll 1$  and 
$\exp[-K_z^2 (Z_{\rm max} - z)^2 ] \ll 1$  
one can approximately write   
\begin{equation}
         A(K_z) \, \approx \,
     \sqrt{{4 \, K_z \, Z_{\rm max} \over
      3 \, \sqrt{\pi} \, {\rm N}_{\rm tot}}} \, .
\label{AKz}
\end{equation}
Such a normalization 
allows to measure naturally 
two-dimensional wavelet coefficients $S(K_z, z)$  
in units of the 
standard deviation. 

To decontaminate 
the wavelet coefficients 
$S(K_z, z)$  from noise
(e.g., \citealt{rhb03, rhb04}) 
at small scales  (high $K_z$)
and from the trend at large scales 
(low $K_z$) we exploit so-called spectral  
filtration.  
For this aim, one can calculate 
the energy spectrum of 
the wavelet transform 
\begin{equation}
        E_S (K_z)  \simeq \, \sum_{n} \, S^2 (K_z, z_n),
\label{ES}
\end{equation}
where index $n=1, 2,... 363$ runs over the chosen 
interval ($0.04 \leq z \leq 3.67$) 
with a small step
($\delta_z=0.01$).

If we admit that  
the distribution of $S(K_z, z)$ is Gaussian
(with zero mean value and unit dispersion)
and all $S(K_z, z_n)$
in Eq.~(\ref{ES}) 
are statistically independent
(see below), then
the value  $E_S (K_z)$ 
obeys to the $\chi^2$--distribution
with 363 degrees of freedom. In this case
we can fix a certain
tabulated value   
$\chi_{1-p}^2$ 
at a given significance level $p$
(e.g., $3 \sigma$, $p=0.0025$)
and select the region of $K_z \leq 28$,  where
$E_S (K_z)$ exceeds this level
(critical interval for the hypothesis 
of the random distribution
of the redshifts). 
On the other hand, at $K_z \la 14$ the value
$E_S (K_z)$  displays a sharp rise caused by
the smooth dependence (trend) 
of the  $z$-distribution
(see Fig.~\ref{AS91-03}).
As a result we choose a representative
interval $15 \leq K_z \leq 28$,
which is  consistent with
significant variations of the z-distribution, and
cancel $S(K_z)$ for all others $K_z$.

\begin{figure*}    
\centering
\includegraphics[width=120mm,  bb=85  200  540  760, clip]{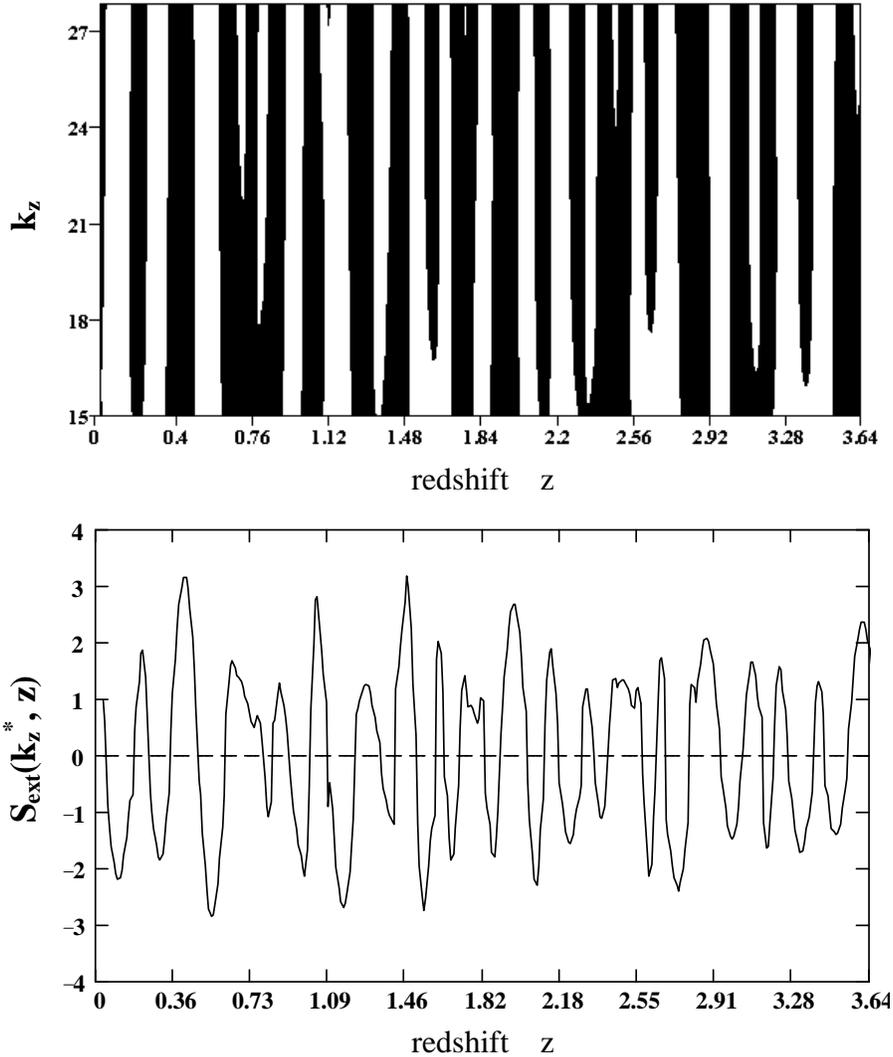}
\vspace{-0.2cm}
\caption{
Results of the wavelet transformation.
{\it Upper panel:} two-dimensional coefficients of the
wavelet transform $S(K_z, z)$;
black strips illustrate positive values ($S >0$),
white strips -- negative values ($S < 0$).
{\it Lower panel:} extremal values of $S(K_z, z)$
satisfying to the condition 
$|S_{\rm ext}(K_z^*, z)| =$ max $(|S(K_z, z)|)$,  
where $K_z$, and $K_z^*$ belong to the intervals 
$15 \leq (K_z,\ K_z^*) \leq 28$.  
}
\label{Wavelet}
\end{figure*}

The upper panel of
Fig.~\ref{Wavelet} 
illustrates the
two-dimensional wavelet coefficients $S(K_z, z)$
in the chosen interval of $K_z$. One can see
almost vertical alternating  black and white strips
corresponding to positive  (enhanced) and negative (reduced)
values of the wavelet coefficients, respectively.
Centers of the strips do not change for different $K_z$
and these centers are consistent with 
the centers of peaks $z_{\rm max}$
and dips $z_{\rm min}$ obtained earlier
in  Papers~I and II (see Table~1). 

\renewcommand{\arraystretch}{1.2}
\begin{table*}   
\caption[]{Redshift values
$z_{\rm max}$ (maxima) and $z_{\rm min}$
(minima) of the 
$z$-distribution of ALSs$^{~a)}$.}
\label{tab:maxmin}
\begin{center}
\begin{tabular}{||c|c|c|c|c|c|c|c|c|c|c||}
\hline
\hline
\multicolumn{1}{||c|}{299 C IV \& 216 Mg II} &
\multicolumn{1}{|c|}{ \, }   &
\multicolumn{4}{|c|}{847 ALSs}  &
\multicolumn{1}{|c|}{ \, }  &
\multicolumn{4}{|c||}{2003 ALSs}   \\
\multicolumn{1}{||c|}{{\it Paper~I}} &
\multicolumn{1}{|c|}{ \, }  &
\multicolumn{4}{|c|}{{\it Junkkarinen et al.~(1991)}} &
\multicolumn{1}{|c|}{ \, }  &
\multicolumn{4}{|c||}{{\it Ryabinkov et al.\ (2003)}}  \\
\multicolumn{1}{||c|}{$N_{\rm AS}(z)$ -- bin statistics }  &
\multicolumn{1}{|c|}{ \, }  &
\multicolumn{4}{|c|}{$N_{\rm AS}(z)$ -- bin statistics }   &
\multicolumn{1}{|c|}{ \, }  & 
\multicolumn{4}{|c||}{Wavelet analysis}  \\
\hline
${\bf z_{\rm max}}$  &     \,           &
${\bf z_{\rm min}}$  &  Significance    &
${\bf z_{\rm max}}$  &  Significance    &   \,   &
${\bf z_{\rm min}}$  &  Significance    &
${\bf z_{\rm max}}$  &  Significance    \\
\,  & \, &  \, & level & \, & level & \, & \, & level & \, & level  \\
\hline
\hline
   \, &  \, & 0.10  & 2$\sigma$ & 0.22 & 2$\sigma$ & \, &
0.11  &   2$\sigma$  & 0.22 & 2$\sigma$  \\
\hline
0.44 & \, & 0.31  & 3$\sigma$ & 0.45 & 3$\sigma$ & \, &
0.30  &   2$\sigma$  & 0.42 & 3$\sigma$  \\
\hline
 \,   &  \, &  \,  &   \,  &  \,  &  \,  & \, &
0.55  & 3$\sigma$  & 0.68 & 1.5$\sigma$   \\
0.77  & \, & 0.58  & 3$\sigma$ & {\bf 0.79}  & 3$\sigma$ & \, &
 --   &   \,  &     --    &   \,         \\
 \,   &   \,  & \,  &  \,   &  \,  & \, & \, &
0.81  &  1$\sigma$   & 0.87 & 1$\sigma$  \\
\hline
1.04  & \, &  0.96 & 2$\sigma$ & 1.10  & 1$\sigma$ & \, &
0.97  &  2$\sigma$   & 1.05 & 3$\sigma$  \\
\hline
1.30  & \, &  1.18 & 1$\sigma$ & 1.28  & 1$\sigma$ & \, &
1.16  &  2.5$\sigma$   & 1.28 & 1$\sigma$ \\
\hline
1.46  & \, &  1.35  & 1$\sigma$ & 1.44 & 2$\sigma$  & \, & 
1.38  & 1$\sigma$    & 1.46 & 3$\sigma$   \\
\hline
1.63  & \, & 1.54  & 2$\sigma$ & 1.63 & 2$\sigma$  & \,  &
1.55  & 2.5$\sigma$    & 1.62 & 2$\sigma$  \\
\hline
1.78  & \, & 1.71  & 1$\sigma$ & 1.75 & 1$\sigma$   & \, &
1.68  & 2$\sigma$    & 1.77 & 1$\sigma$    \\
\hline
1.98 & \,  & 1.84  & 1$\sigma$ & 1.96 & 2$\sigma$   & \, &
1.87  & 2$\sigma$  & 1.97 & 2.5$\sigma$     \\
\hline
2.14  & \, & 2.07  & 1$\sigma$ & 2.13 & 1$\sigma$   & \, & 
2.07  & 2$\sigma$    & 2.15 & 2$\sigma$     \\
\hline
 --  & \,  &  2.24 & 2$\sigma$ & -- &  \,  & \, &
2.23  & 1.5$\sigma$    & {\bf 2.31} & 1$\sigma$     \\
2.45 & \,  & --  & \,     &  2.42   & 3$\sigma$ & \, &
2.38  & 1$\sigma$  & 2.49 & 1.5$\sigma$     \\
\hline
2.64  & \,  &  --   &  \,       & --  &  \,  & \, &
2.61  & 2$\sigma$  & {\bf 2.66} & 1.5$\sigma$     \\
\hline
2.86  & \, & 2.72  & 3$\sigma$ & 2.87 & 3$\sigma$   & \, &
2.74  & 2$\sigma$    & 2.87 & 2$\sigma$     \\
\hline
 \,   &  \, & \,   &   \,       & \,  &  \,   & \,    &
3.00  & 1.5$\sigma$  & 3.09  & 1.5$\sigma$   \\
 --   & \, & 3.05  & 2$\sigma$ & {\bf 3.16} & 2$\sigma$  & \, &
 --   &  \,       &   --    & \,            \\
 \,   & \,  &   \,  &   \,  &  \, &  \,  & \, &
3.16  & 1.5$\sigma$  & 3.22 & 1.5$\sigma$    \\
\hline
 --   & \, & 3.40   &  3$\sigma$  & --  & \,   & \, &
3.32  & 1.5$\sigma$    & {\bf 3.41} & 1$\sigma$    \\
 --   & \, & --  & \,   & 3.56   & 2$\sigma$  & \, &
3.49  & 1.5$\sigma$  & 3.61 & 2.5$\sigma$    \\
\hline
\hline
\end{tabular}
\end{center}

\medskip  
$^{a)}$ Bold values $z_{\rm max}$  mark divergences between our 
previous and present results  \\
\end{table*}
\renewcommand{\arraystretch}{1.0}

The lower panel of  Fig.~\ref{Wavelet}  illustrates
how wavelet coefficients allow one
to localize positions of the centers
of the  bands
and estimate significance levels of the
peaks and dips.  
The panel shows the z-dependence of the 
extremal wavelet coefficients
$S_{\rm ext}(K_z^*, z) \equiv S(K_z^*, z)$,
where  the values $K_z^*=K_z^*(z)$  
are obtained numerically 
for each fixed $z$
from the condition
$|S_{\rm ext}(K_z^*, z)|=$
max($|S(K_z,  z)|$).
The extremal wavelet coefficients 
represent the variations of
$S(K_z, z)$ in the most 
pronounced form.\footnote{Note that, 
the results are not too
sensitive to the form of presentation.
Thus, fixing a certain value
$K_z$ within the chosen interval
(e.g., $K_z=20$) one can obtain the curve
$S(20, z)$ which is weakly different from
the curve shown in Fig.~\ref{Wavelet}}. 

Positions of peaks $z_{\rm max}$ and
dips $z_{\rm min}$ 
are determined as weighted
mean values (centers of gravity) of the points
$S_{\rm ext}(K_z^*, z)$ forming the peaks
or dips  with respect to zero level.
The accuracy of such determination is $\pm 0.02$.
As it is seen from Table~1  majority of the values
$z_{\rm max}$ and $z_{\rm min}$  calculated
for the wavelet coefficients are close
to the values
$z_{\rm max}$ and $z_{\rm min}$ obtained
in our previous statistical studies. 
There are only a few exceptions
(indicated in Table~1 by bold font) 
concerned with two cases of splitting of 
peaks into double ones at $z_{\rm max}= 0.79$ and
3.16, and  three cases of appearance
of new peaks at  $z_{\rm max}= 2.31$, 2.66 and
3.41. The latter peaks are located 
on the decreasing part of the $N_{\rm AS}(z)$
distribution  (Fig.~\ref{AS91-03})
where statistics is essentially poorer.
Note that the peak at $z_{\rm max}= 2.66 \pm 0.02$
was obtained earlier
by the statistical analysis of C~IV doublets
(first column in  Table~1).

It follows from Table~1
that, in general,  there is  
a good agreement between
the new  and previous results.
In addition,
the significance levels of the peaks and dips  
estimated using
the wavelet coefficients
$S_{\rm ext}(K_z^*, z)$
are systematically 
higher than similar significance
levels obtained in  Papers~I and II.
\begin{figure*}    
\centering
\includegraphics[width=120mm,  bb=85  95  540  740, clip]{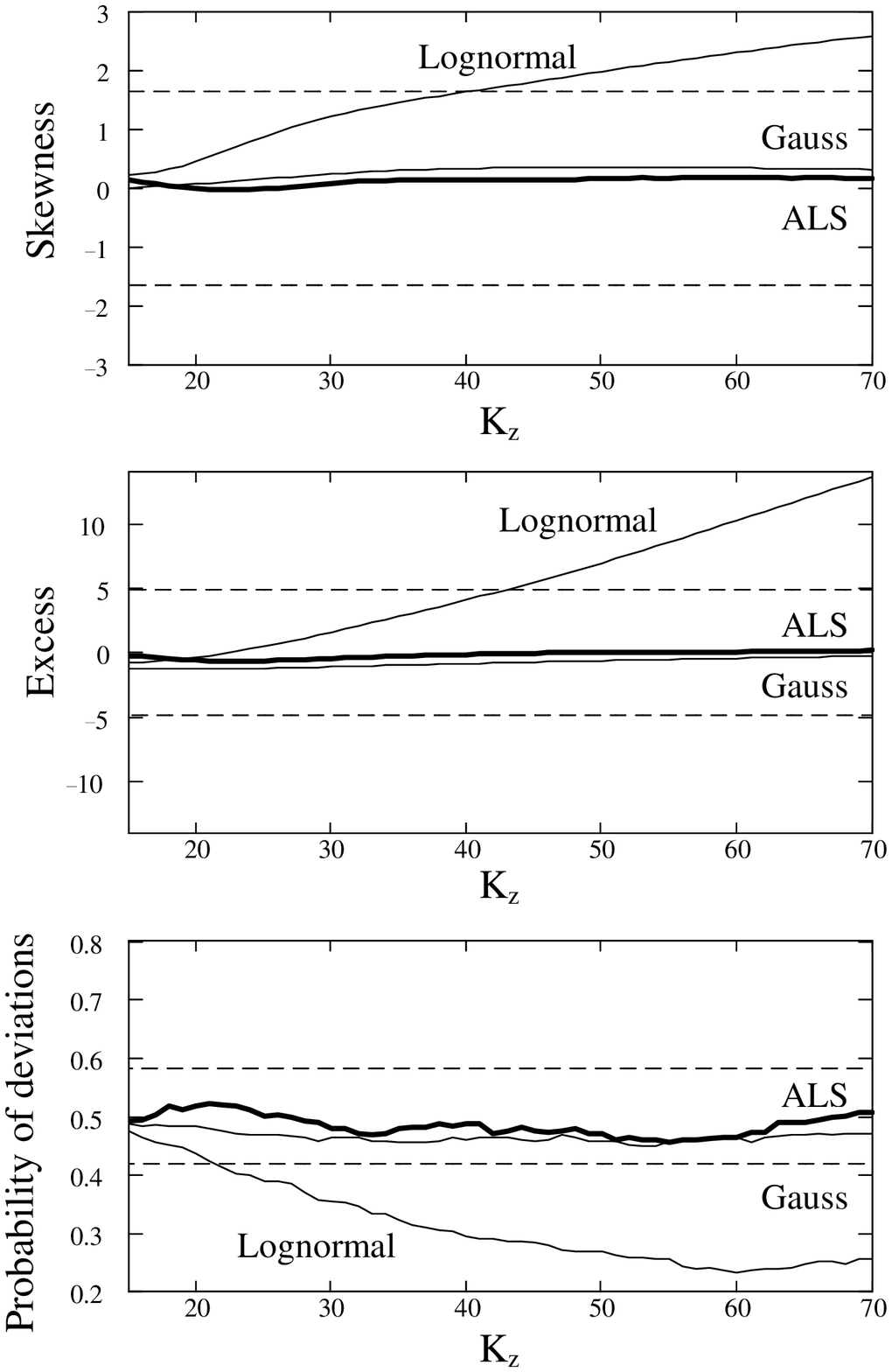}
\vspace{-0.2cm}
\caption{
Statistical comparison of wavelet transformations
produced for the $z$-distribution of ALSs (thick lines),
the Gaussian distribution of 2000 random $z$-points
with periodical variations of the mean value 
(thin lines -- `Gauss'), 
the lognormal distribution
simulated on the base of the same Gaussian model
(thin lines -- `Lognormal'; see text).     
{\it Upper panel:} skewness, {\it middle panel:}
excess (kurtosis), {\it lower panel:} probability 
of $\pm$-deviations in a dependence on 
$K_z$ within the interval
$15 \leq K_z \leq 70$.   
Dashed lines indicate 
the boundaries of critical regions at
the significance level $3 \sigma$.
}
\label{Lognorm}
\end{figure*}

A set of our numerical simulations of the 
wavelet coefficients $S(K_z, z)$
produced    
for Poisson distribution of $N_{\rm tot}$ points 
$z_i$  using   
Eqs.~(\ref{S(z)}) and (\ref{AKz})
has shown that the mean squared deviation 
$\sigma$, or the variance $\sigma^2$,
calculated for the wavelet coefficients
turned out  to be lower than 
unity.  Actually, the 
random wavelet coefficients obtained 
at different $z$ and $K_z$  
are not independent. In particular, 
we have $\sigma^2[S(K_z, z)] = 0.56$ 
at $K_z=15$ and  $\sigma^2[S(K_z, z)] = 0.66$ 
at $K_z=28$. Such a decrease 
of the variance results in 
an increase of  
the significance levels of $S(K_z, z)$.   
Consequently,    
the significance levels of peaks and dips 
indicated in  Table~1 may be regarded as 
minimal estimations.    
Additional calculations have shown  
that statistical dependence of $S(K_z, z)$ 
at different $K_z$ and $z$ in 
Eq.~(\ref{ES}) leads to 
lowering of the boundary 
$\chi_{1-p}^2$ and  as a result 
to expansion of 
the critical interval of $K_z$ 
into the range of higher $K_z$.
Thus the chosen 
interval $15 \leq K_z \leq 28$ may be treated as
minimal critical interval 
or minimal representative interval 
appropriate to the alternative hypothesis
of a regular origin of the peaks and dips 
in Fig. \ref{Wavelet}.      
   
In this section and hereafter we use 
the hypothesis of Gaussian
distribution of the absorbing matter as  basic one 
for estimations of the statistical significance
of the peaks and dips and their regularity. 
It follows from the results discussed  
that the simple Gaussian model for  
the  distribution of ALSs   
may be rejected on the 
significance level $\ga  2\sigma$. 
However, it is worthwhile to verify
consistency of the ALS distribution
with a wider class of   
Gaussian-like and non-Gaussian
distributions.  
For this aim, we 
simulated two particular
stochastic models  
(basic hypotheses)   
of the absorbing
matter distribution:  \\ 
(i) {\it Gaussian binned distribution}
of 2000 redshift points $z_i$ 
with a periodical expectation $\mu(z)$
and the variance
equal to unity:
\begin{equation}
     P\left[ x(z) \right] = {1 \over \sqrt{2 \pi}} \exp 
     \left\{ - \left[ x(z) - \mu(z) \right]^2/2 \right\},    
\label{Gauss}
\end{equation}
where 
$x(z)=[n(z)-n_0]/\sqrt{n_0}$, $\ n(z)$ 
is a number of 
random points within a bin $z \pm \Delta_{\rm z}/2,\, $
$\Delta_{\rm z} = 0.071$,
$n_0$ is a mean number of points
averaged over all bins;   
$\mu(z)=2.5 \cos(2 \pi z\ 16/Z_{\rm max})$,
numbers 2.5 and 16 are
chosen as
an amplitude (significance level) 
and harmonic number
of  the variations;\
(ii) so-called {\it lognormal random field} 
suggested in literature 
as a plausible model  for 
density fluctuations in the Universe 
(e.g., \citealt{cj91}, \citealt{bd97}).  
The lognormal distribution of 
the same number of points
has been  produced  
numerically 
from the distribution (\ref{Gauss})
by a transformation of the variable
$x(z)$  into $y(z)=\exp [x(z)]$.
Accordingly, 
the  wavelet transforms $S(K_z, z)$
have been calculated
for both distributions
using simple modification of Eq.~(\ref{S(z)})
adjusted to binned statistics.

Fig.~\ref{Lognorm} displays 
a comparison of 
statistical symmetry properties of 
the wavelet transformations 
produced for two models (i) and (ii)
with  $S(K_z, z)$  calculated for ALSs.
Three panels in the Fig.~\ref{Lognorm}
represent (from top to bottom)
the skewness, kurtosis (excess),
and probability of sign deviations 
(criterion of signs) 
calculated in standard technique.    
One can notice that the symmetry properties
of the wavelet coefficients obtained for
ALSs and `Gauss' model are statistically
close. For both samples
the skewness and excess  
are situated near 
zero for all  values $K_z$
and the probability of $\pm$ deviations 
variates near the number 0.5  
corresponding to
symmetrical distributions. 
On the other hand, the `Lognormal'
model  notably differs from 
both symmetrical distributions
by its asymmetry.  

These results and results of 
our additional calculations 
carried out for the standard Gaussian
field and its transformation into
the lognormal random field  
confirm that
the Gaussian (or Gaussian-like) 
distribution may be chosen 
as a background for the statistical
analysis of ALS.    
Let us remind as well that 
in our consideration
we treat  all ALSs 
detected along each line of sight 
and fallen into
the averaging velocity interval $\delta v$ 
(predominantly  $\delta v=500$~km/s)
as a single redshift.
That smoothes away all nonlinear stochastic
fluctuations of the $z$-distribution  
and  approaches
them to Gaussian-like random field. 

\section{Isotropy of the distributions}
\label{isotr}
Let us test isotropy (or anisotropy) of the ALS distribution
comparing the z-dependences of the wavelet coefficients
$S_{\rm ext}(K_z^*, z)$
calculated for different hemispheres.
The procedure is similar to that used in
Papers~I and II 
for verification  
of a dependence
of $N_{\rm AS}(z)$ (see Section~1)
on  orientations of hemispheres.

\begin{figure}
\vspace{-0.2cm}
\centering
\includegraphics[width=120mm,  bb=65  175  510  745, clip]{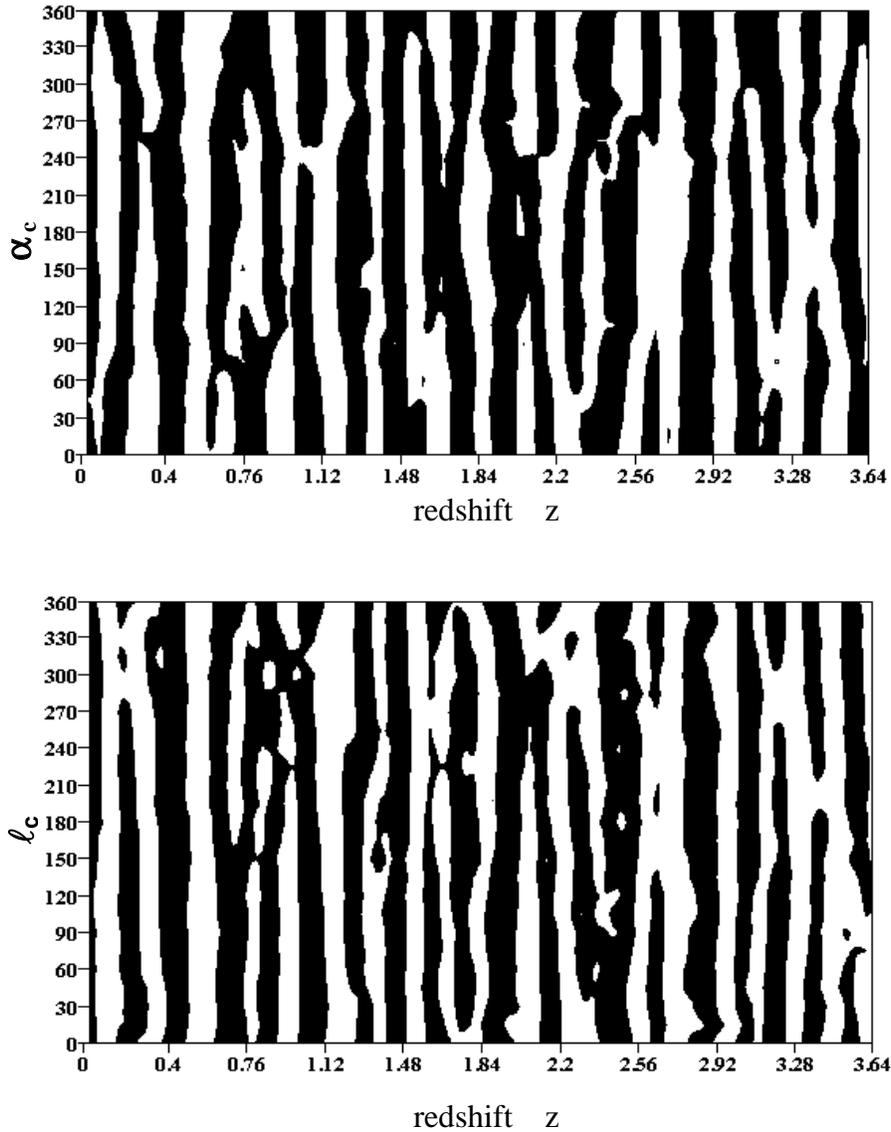}
\caption{
$z$-dependences of the wavelet coefficients
$S_{\rm ext}(K_z^*, z)$ (see text) calculated
for 24 hemispheres in the equatorial ({\it upper panel}) 
and Galactic ({\it lower panel}) coordinate systems. 
{\it Upper panel}: 
the hemispheres are determined by 
the right ascensions 
$\alpha_c - 90^\circ \leq \alpha \leq  \alpha_c + 90^\circ$,
$\alpha_c$ stands for hemisphere centers,  
and declinations $-90^\circ \leq \delta \leq +90^\circ$;
the vertical axis represents 
$\alpha_c$ in degrees plotted 
with the step $15^\circ$ ($1^h$). 
{\it Lower panel}:
the hemispheres are determined by 
the Galactic longitudes 
$l_c - 90^\circ \leq l \leq  l_c + 90^\circ$,
$l_c$ stands for hemisphere centers, 
and latitudes $-90^\circ \leq b \leq +90^\circ$;
the vertical axis represents $l_c$ in degrees
plotted with the step $15^\circ$.}
\label{Strips}
\end{figure}

Fig.~\ref{Strips} shows two sets of redshift
dependences of the value $S_{\rm ext}(K_z^*, z)$
plotted for 24 hemispheres. 
Both panels in Fig.~\ref{Strips}  
may be treated as two-dimensional distributions
of the wavelet coefficients $S_{\rm ext}$.
The vertical axis on the upper panel (equatorial CS)
represents the right 
ascension of the hemisphere centers $\alpha_c$ (in degrees),
the horizontal axes corresponds to the redshifts $z$. The black
strips display positive (enhanced) values of
the wavelet coefficients, $S_{\rm ext} >0$,
and the white strips display negative (reduced) values
of the wavelet coefficients, $S_{\rm ext} < 0$.
One can see the set of continuous vertical black and white
strips with centers at the same values of $z_{\rm max}$
and $z_{\rm min}$ 
as indicated in Table~1, i.e.,
$z_{\rm max}=$ 0.42, 1.05, 1.28, 1.46, 1.97, 2.87, 3.09
and $z_{\rm min}=$ 0.11, 0.30, 0.55, 0.97, 1.16, 1.38,
1.87, 2.23, 2.74, 3.00, 3.32.
These values of $z_{\rm max}$ and $z_{\rm min}$
are independent of $\alpha_c$ within statistical errors. 
The rest black and white strips are more fragmentary.
However, the strips comprise segments which 
are notably longer than $180^\circ$.
Such segments
exceed possible effects of   
enhanced concentration of ALSs (clumps) at certain $z$
in separate directions and
can not be
reproduced  by 
overlapping of hemispheres.

The vertical axis on the lower panel represents the Galactic
longitude of the hemisphere centers $l_c$ (in degrees).
It is clear from comparison of the upper and lower panels
that the positions of the vertical strips as well as their
width do not change (within statistical uncertainties) 
if we use the Galactic
coordinates instead of the equatorial ones. 
Consequently, the wavelet coefficients
do not depend substantially on the orientation of
hemispheres. 
We calculated  the correlation
coefficients for the $z$-distributions in 12 pairs
of opposite (independent) hemispheres and found
moderate positive correlation 
at the significance level 
$\ga 2 - 3 \sigma$,
probably due to 
variations of peak-and-dip
profiles with relatively
stable centers of gravity. 
 
On the other hand, our calculations of  
the wavelet coefficients    
within separate
quadrants ($\alpha_c \pm 45^\circ$)  
instead of hemispheres
($\alpha_c \pm 90^\circ$)  
display more irregular
black and white 
domains around certain $z_{\rm max}$ and $z_{\rm min}$. 
As a result the strips-like
picture (as in Fig.~\ref{Strips}) 
turns out to be more fragmentary.

Thus our wavelet analysis of 
the ALS redshifts 
gives  evidence in favour of
(i) the existence of the statistically significant 
($\ga 2 \sigma$) peaks and dips
in the z-distribution of the ALSs at certain 
$z_{\rm max}$  and  $z_{\rm min}$  and
(ii) approximate independence 
$z_{\rm max}$ and $z_{\rm min}$
obtained for different hemispheres
of  hemisphere orientations.

\section{Regularity (quasi-periodicity) of the peaks and dips}
\label{period}
The positions of the maxima 
$z_{\rm max}$  and  minima $z_{\rm min}$
presented in Table~1 
are distributed nonuniformly.
For instance, it may be 
verified  numerically 
that the power spectrum 
calculated for the whole sequence of redshift points
(see below) displays no significant peaks 
in a wide region of  
$z$-scales ($z$-periods) 
including all  
intervals between neighbour maxima  
or  minima in Table~1. 
Therefore, to reveal some regularity
(e.g., quasi-periodicity) in the sequence of the peaks and dips
we need to use so-called rescaling 
functions $\tau=f(z)$ which satisfy,
e.g., to periodicity condition:
\begin{equation}
   \tau_{\rm max}^{\rm m+1} - \tau_{\rm min}^{\rm m} =
    f(z_{\rm max}^{\rm m+1}) - f(z_{\rm min}^{\rm m}) = {\rm const},
\label{tau-tau}
\end{equation}
where $m=1, 2, ...$ -- is the continuous 
numeration of the maxima and minima.
 
In  Papers~I and II
we tested the quasi-periodicity  of the ALS
distribution  using several simple rescaling (trial) 
functions of $z$ 
including the function $\log (1+z)$
which  attracts special attention 
in literature in search of a 
periodicity  
of the QSO $z$-distribution (see Section~6).
In contrast to  these papers, our new 
calculations based on the extended statistical sample
have not revealed a significant (at level $> 2\sigma$)
periodicity with respect 
to the same trial functions.
Therefore,  
we perform the power-spectrum analysis
using  some special trial function
(unless the different functions are discussed) 
which appears to be more appropriate for testing
of the quasi-periodicity of the ALS distribution:
\begin{equation}
     \tau_i = \tau(z_i) = \sinh^{-1} [(\eta(z_i)-a_1)\,  a_2],
\label{tau}
\end{equation}
where $i=1,2,... {\rm N}_{\rm tot}$ is a numeration of all ALSs 
(N$_{\rm tot}=2003$); $a_1$ and $a_2$ are constants;
$\eta(z)$ is the {\it conformal time} defined by
\begin{eqnarray}
  \eta(z_i) & =  & \int_0^{z_i} 
      {1 \over \sqrt{\Omega_{\rm M} (1+z)^3 +
      (1-\Omega_{\rm M} - \Omega_{\rm \Lambda})(1+z)^2 +
       \Omega_{\rm \Lambda}}} \, \,  {\rm d}z ,
\nonumber   \\
    \, &  \, & \,
\label{eta}
\end{eqnarray}
where we choose $\Omega_{\rm M}=0.3$ and
$\Omega_{\rm \Lambda}=0.7$.

\begin{figure*}     
\vspace{-0.2cm}
\centering
\includegraphics[width=120mm,  bb=80  135  520  735,  clip]{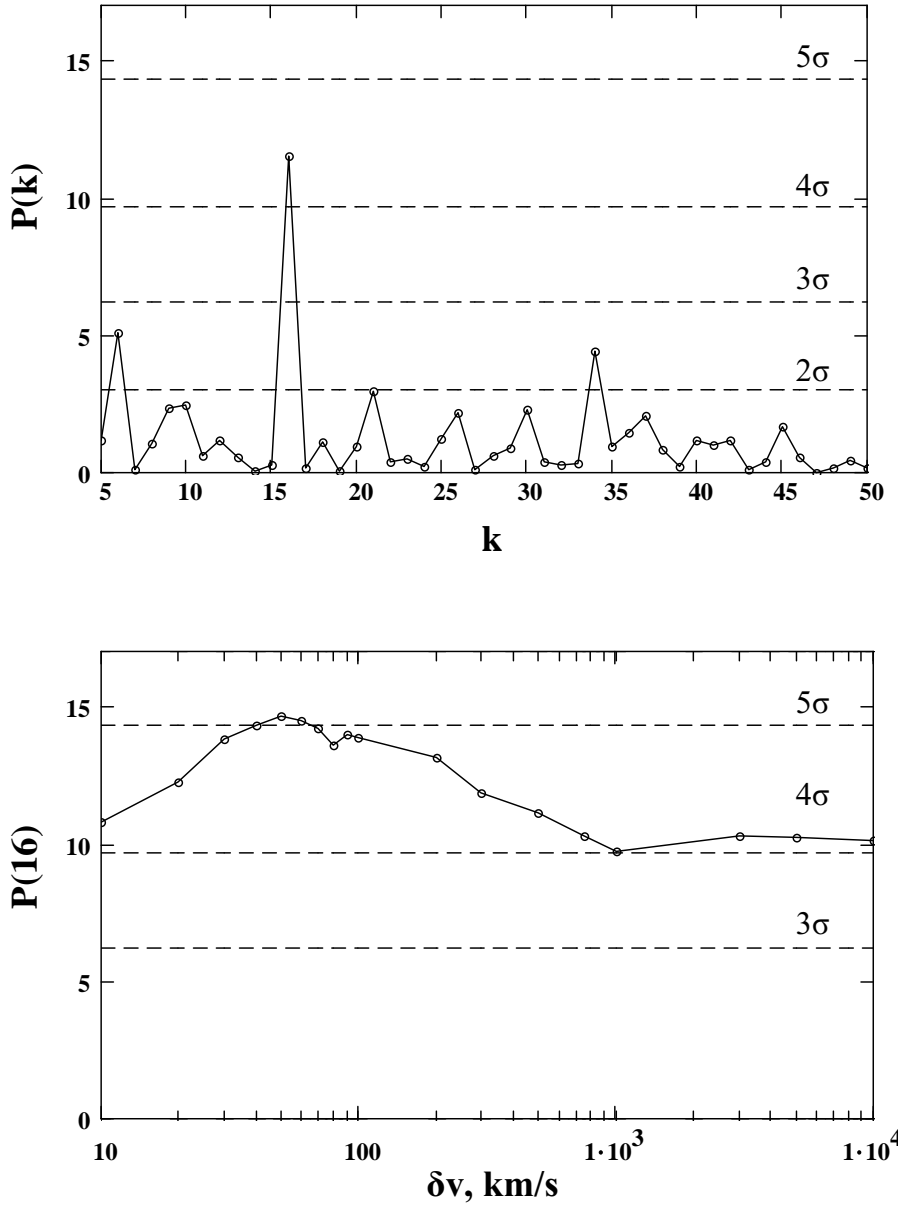}
\caption{
Power spectra  P$(k)$,  $k$ is the harmonic number,
calculated with using  the trial
function  \protect{(\ref{tau}),\  (\ref{eta})}
for the complete sample of the values 
$z_i,\,  i=1, 2, ... {\rm N}_{\rm tot}$ 
(see text).
{\it Upper panel} represents  P$(k)$
defined by
Eq.\ (\protect{\ref{P_k}})
at the velocity interval
$\delta v = 500$km~s$^{-1}$
of AS associations
(along each line of sight)
into single redshifts $z_i$ (see Section\ 1).
The main peak corresponds to $k=16$.
{\it Lower panel} plots the dependence  
of the highest-peak value P$(k=16)$ on 
the interval $\delta v$.
The horizontal dash lines
specify the confidence levels  
$\beta = 0.95$ 
($2\sigma$; 
only on the upper panel),
$\beta = 0.998\ (3\sigma)$, $\beta = 0.999936\ (4\sigma)$,
$\beta = 0.9999994\ (5\sigma)$, respectively.}
\label{P(k)}
\end{figure*}

The spectral power 
of the distribution
is calculated without the use of statistical
bins $\Delta_{\rm z}$ 
(e.g.,  \citealt{k77},  \citealt{arpetal90},  Paper~I):
%
%
%
\begin{equation}
          {\rm P}(k)  =   {1 \over {\rm N}_{\rm tot}}
                          \left\{ \left[
          \sum_{i=1}^{{\rm N}_{\rm tot}} \cos \left(
          {2\pi k \tau_i  \over \hat{\tau}}
              \right) \right]^2  \, +  \, 
          \left[ \sum_{i=1}^{{\rm N}_{\rm tot}}
             \sin \left(
       {2\pi k \tau_i  \over \hat{\tau}}
              \right) \right]^2 \right\},
\label{P_k}
\end{equation}
where 
$\tau_i$ is given by Eq.~(\ref{tau}),
$\hat{\tau}$
$=\tau (z_{{\rm N}_{\rm tot}}) - \tau (z_1)$
is the whole interval of values $\tau_i$, and
$k$ is a harmonic number.
The periodicity  yields the peak in the
power spectrum ${\cal P}= {\rm max}$ (P$(k)$) with a
confidence probability
\begin{equation}
   \beta = [1- \exp(-{\cal P})],
\label{beta}
\end{equation}
where the confidence level is defined with respect
to the hypothesis of the uniform distribution
of the values  $\tau_i$.

Results of the power-spectrum analysis
are sensitive to 
the averaging velocity interval $\delta v$
of the ALS associations 
into separate redshifts (see Sect.\ \ref{intro}).
The upper panel of Fig.\ \ref{P(k)}
plots the results of the calculations
based on Eq.\ (\ref{P_k}), where 
$\delta v$=500~km~s$^{-1}$ and
the constants in Eq.\ (\ref{tau}) 
are $a_1$=1.26,  $a_2$=3.37.   
One can see that 
in the represented interval of $k$
the only appreciable peak P$(k)$ appears
for $k=16$  at the significance level
$> 4\sigma$. 
This peak corresponds to a periodicity
in units of the function
(\ref{tau}) with the period
$\Delta \tau =\hat{\tau}/k = 0.199\pm 0.001$.
Note that the most significant peak  
is very sensitive to variations of
the values  $a_1$ and $a_2$ at fixed
$\Omega_{\rm M}$ and $\Omega_{\rm \Lambda}$
(or variations of $\Omega_{\rm M}$ and $\Omega_{\rm \Lambda}$
at fixed $a_1$ and $a_2$). 
The deviations  of $a_1$ 
and $a_2$
from indicated values
exceeding $\pm 0.01$ and $\pm 0.28$, 
respectively, 
reduce the peak P$(k)$ to the   
significance level $\la 3 \sigma$.

The lower panel of Fig.~\ref{P(k)} 
shows the dependence
of the main peak amplitude P$(k=16)$
on the velocity interval $\delta v$.
It is seen that the maximum of P$(k=16)$
increases smoothly with decreasing of $\delta v$
and reaches the significance level 
$> 5\sigma$ at 
$\delta v=$50--70~km~s$^{-1}$,  
being basically robust to
variations of  $\delta v$. 
Let us note that such an averaging 
of redshifts
along each line of sight 
(up to $\delta v =10,000$~km/s)
smoothes down 
possible effects of {\it longitudinal} clumping
of absorbing matter.  

\begin{figure*}     
\vspace{-0.2cm}
\centering
\includegraphics[width=120mm,  bb=80  440  540  740, clip]{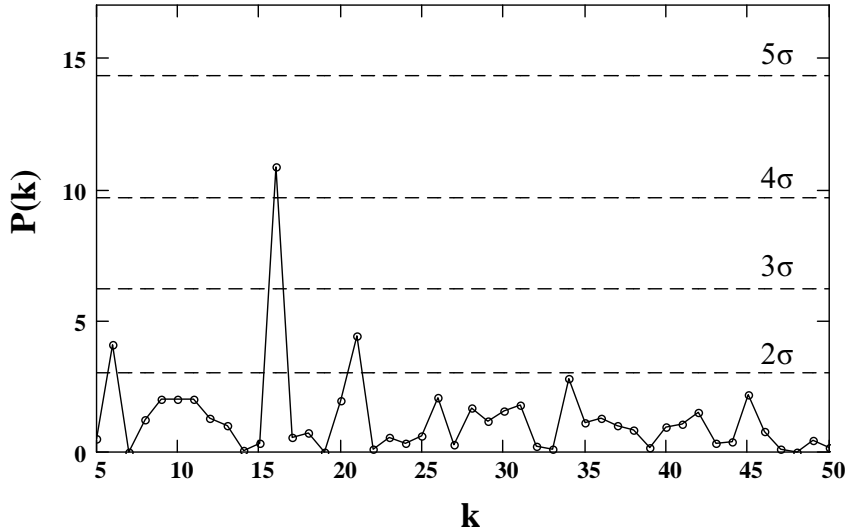}
\caption{
Same as in upper panel of Fig.~\ref{P(k)}
but for 1778 ALSs  sampled with elimination of  
all groups of  
QSOs with their ALSs 
located 
within $1 \times 1$~deg$^2$
square regions on the sky.
}
\label{P(k)-clumps}
\end{figure*}

In addition Fig.~\ref{P(k)-clumps} 
demonstrates  effects of 
plausible clumping of ALSs in fields perpendicular to
lines of sight. We calculate 
the same power-spectrum 
as in Fig.~\ref{P(k)} but with
elimination of   
all groups of QSO spectra
if their coordinates turn out to locate 
within any $1 \times 1$~deg$^2$
square region on the celestial sphere.   
Similar  significant peak $P(k)$ 
at $k=16$ as in  Fig.~\ref{P(k)} 
indicates  weakness of 
{\it transversal} clumping effects 
in our analysis.  

The rescaling function given by Eqs. ~(\ref{tau}) and (\ref{eta})  
is not unique. For example,  
following  \citet{asss04} 
we may choose a
model independent ansatz for
the Hubble parameter 
\begin{eqnarray}
         H(z) & = & H_0 \, 
	 \sqrt{\Omega_{{\rm M}0} \, (1+z)^3 + A_0 + A_1 (1+z) + A_2\, (1+z)^2};  
\nonumber    \\	    
         \Omega_{{\rm M}0} & = & 8 \pi G\ \rho_0/(3  H_0^2),
\label{H}
\end{eqnarray}
where $H_0$ and  $\rho_0$ are the Hubble constant 
and matter density   
at the present epoch ($z = 0$).   
We calculate $P(k)$  
replacing $\tau_i$ and $\hat{\tau}$
in Eq.\ (\ref{P_k})
by  the conformal time  $\eta_i$  and  $\hat{\eta}$
$=\eta (z_{{\rm N}_{\rm tot}}) - \eta (z_1)$,
where the rescaling function
$\eta(z)$ is  given by
Eq.\ (\ref{eta}) with the denominator
replaced  by $H(z)/H_0$
from  (\ref{H}).  
Varying $\Omega_{{\rm M}0}$ and 
the coefficients $A_0, A_1$, $A_2$, 
and keeping the equality 
$A_0+A_1+A_2=1-\Omega_{{\rm M}0}$, we also
obtain the significant peak $P(k)$ 
of the power spectra
at $\Omega_{{\rm M}0}=0.3$  and
$A_0=-0.61, A_1=3.11, A_2=-1.8$
on the level $\ga 4 \sigma$
(slightly lower than that in Fig.\ \ref{P(k)}). 

The other example of a rescaling function  
is the mean comoving 
number density of absorbers associated with filaments
(the elements of so-called large scale structure;
see Sect.\ \ref{concl})  
introduced by \citet*{ddt00} in their Eqs.\  (2.2) and (2.15) 
with the function $B(z)$  
given by their Eq.\ (2.3a).
Employing this rescaling function  
we obtain the peak of $P(k)$ at the same $k=16$
but even more 
significant ($> 5 \sigma$) 
than the peak in Fig.\ \ref{P(k)}.     
However, additional powerful peaks of $P(k)$ 
(less significant than the main peak) 
also appear for this trial function. 
In this case the periodicity
becomes more complex and requires 
a special consideration.   

For comparison,  
using  Eqs.\  (\ref{P_k}),  (\ref{tau}) and (\ref{eta})
we have calculated  additionally
the power-spectrum  $P(k)$
of the  distribution of
emission-line redshifts $z_e$   
detected in the spectra of 
661  original QSOs  
within the interval $z=0.29$--4.9. 
We have not revealed  
significant peaks (with significance 
$\ga 3 \sigma$) for a wide region
of periods around $\Delta \tau \sim 0.2$.
Thus we conclude that 
there is no  straightforward
coupling between possible periodicities 
of the  $z$-distributions of
ALSs and original QSOs.
Note, however, that 
the sample of QSOs 
is not statistically
representative.  

To sum up the results of this
section, we conclude that  
the appearance of 
significant peaks 
in the power spectra 
(calculated with 
different trial functions)
may be regarded as
additional evidence for reality 
of the peak-depression sequence 
in the $z$-distribution of ALSs
and its nonrandom origin. 
At the same time 
this quasi-periodical
distribution does not correlate 
with the distribution 
of the original QSOs.     

\section{One-dimensional correlation function}
\label{CF}
We determine a two-point 
correlation function $\xi(\delta \tau)$
for the sample of ALSs 
in an unconventional way. 
Specifically,
the variable $\delta \tau$
substitutes a 
comoving distance $r$
between pairs of objects
(e.g., galaxies or clusters of galaxies)
in the standard two-point
correlation function $\xi(r)$
(e.g., \citealt{p93}, \citealt{ls93}, 
\citealt*{kss00}, \citealt{jmst04},
and references therein). 
In our consideration $\delta \tau$ is a measure
of the separation of two arbitrary ALSs
numerated by $i$ and $j$ 
($i,j= 1,2,...{\rm N}_{\rm tot}$, $i \neq j$)
in units of  the trial function  
$\delta \tau_{i, j} = |\tau_i - \tau _j | $
introduced by Eq.\  (\ref{tau}). 
Accordingly,
the intervals of the conformal time
given by  (\ref{eta})
and the redshifts are
$\delta \eta_{i, j} = |\eta_i - \eta_j|$
and
$\delta z_{i, j} = |z_i - z_j|$,
respectively.
Thus we can write:
\begin{equation}
     \xi(\delta \tau) = {{\cal N}_{\rm obs}(\delta \tau) \over
                           {\cal N}_{\rm sim}(\delta \tau)} - 1,
\label{xi}
\end{equation}
${\cal N}_{\rm obs}(\delta \tau)$
is the number of observed pairs of ALSs  
separated by the variables $\delta \tau_{i, j}$ 
within the range $\delta \tau \pm \Delta_{\tau}/2$,
where $\Delta_{\tau}=0.07$ is the chosen bin width.
${\cal N}_{\rm sim}(\delta \tau)$ is the calculated
number of cross pairs between the real sample of ASs 
and the points of a random
sample simulated for the same 
intervals of $z$  or $\tau(z)$  and
the same smoothed 
function (trend) as the real 
sample (e.g., \citealt*{mjb92}). 
The both samples  (real and simulated ones)
have the same number of redshift points.

\begin{figure*}     
\vspace{-0.2cm}
\centering
\includegraphics[width=120mm,  bb=75  165  520  770,  clip]{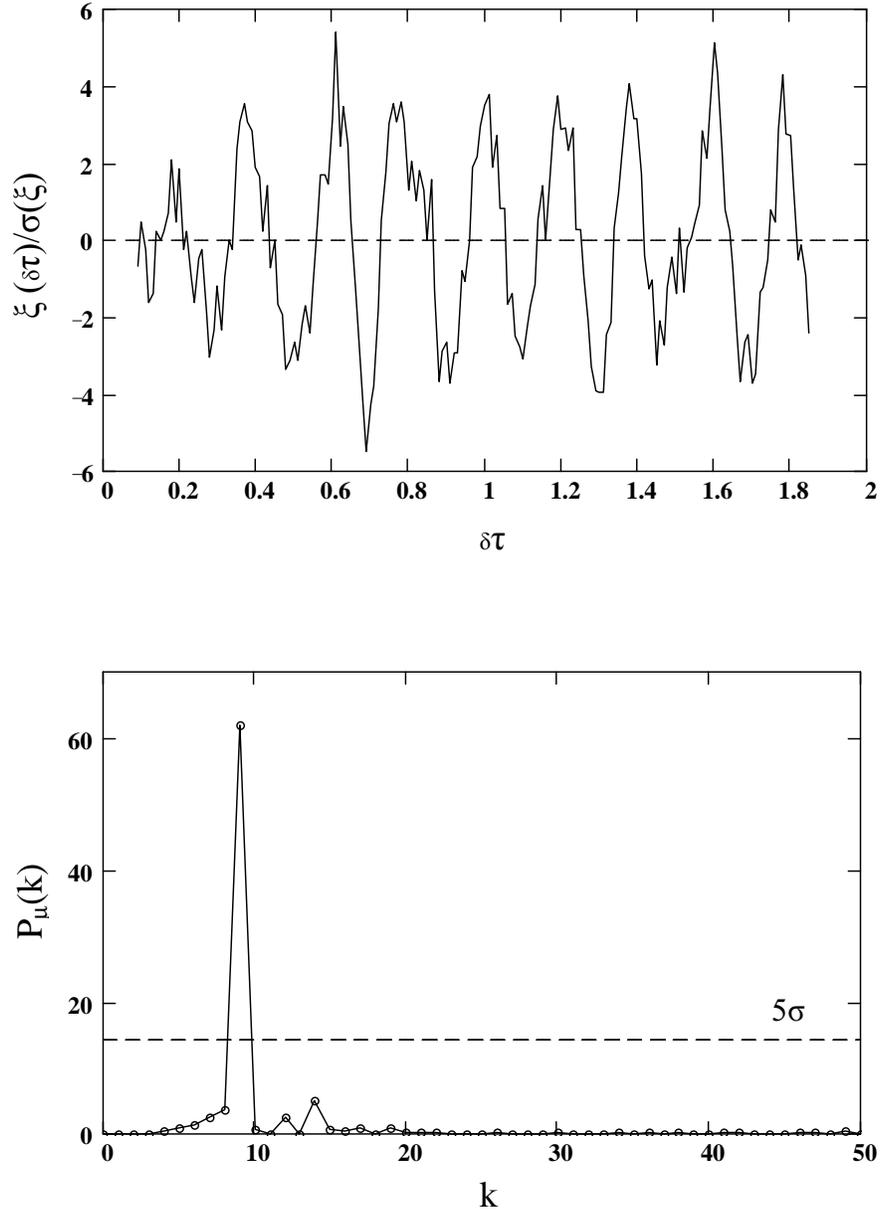}
\caption{
{\it Upper panel}:
one-dimensional correlation function 
$\xi (\delta \tau)/\sigma(\xi)$
(see text) versus
the interval $\delta \tau$  ($0.1 \la \delta \tau \la 1.9$)
between the components of pairs of  ALS redshifts 
(sampled at $\delta v=500$km~s$^{-1}$);
$\tau(z)$ is the trial function given by 
Eq.\ \protect{(\ref{tau})}. 
{\it Lower panel}:  
power spectrum  P$_\mu (k)$ 
calculated  for
$\mu=\mu(\tau)=\xi (\delta \tau)/\sigma(\xi)$
according to 
Eq.\ \protect{(\ref{Pk_mu})}; 
$k$ is the harmonic number.} 
\label{XiSig}
\end{figure*}
Note that  
the variable $\delta \tau$
in Eq.\ (\ref{xi}) may be considered
as a measure of the temporal separation of
any two epochs in units of 
the scaling function  (\ref{tau}). 
Actually, in our case the temporal
interpretation of  $\xi(\delta \tau)$
is more appropriate than the spatial one. 
Unlike  the standard approach to the
spatial two-point correlation functions
(see below) the function (\ref{xi})
is based on the conception of a
single reference center $\tau=0$ ($z=0$)  
concerned with the observer.
Accordingly, 
all sampled  points (redshifts)   $\tau_i(z_i)$ are   
distributed over concentric spherical layers (bins)   
independently of the directions 
to their original QSOs.
We count up all redshifts $\tau_j$ 
inside a concentric
layer ($\tau \pm \Delta_\tau/2$)
with respect to any chosen point $\tau_i$
(out of this layer). All redshifts
within a layer are treated as equivalent    
despite of various spatial (comoving) distances 
between them.   

On the other hand,
the function (\ref{xi}) differs
from the correlation functions calculated
by counting only  line-of-sight pairs of
the objects 
with different mutual comoving distances  
(e.g.,  \citealt*{qby96}, \citealt{bj00}), 
although, the results
of both approaches
may turn out to be quite consistent.

The upper panel of Fig.\ \ref{XiSig}
displays the
one-dimensional (two-point)
correlation function  (\ref{xi})
in units of the appropriate Poisson error
$\sigma(\xi)$  calculated  as follows  
(e.g., \citealt{pn91}):
\begin{equation}
           {\xi(\delta \tau) \over \sigma(\xi)}
         \approx {{\cal N}_{\rm obs} (\delta \tau) -
            {\cal N}_{\rm sim} (\delta \tau)  \over
        \sqrt{{\cal N}_{\rm sim} (\delta \tau) }},
\label{xitosig}
\end{equation}
where
\begin{equation}
           \sigma(\xi) = {1+\xi \over \sqrt{{\cal N}_{\rm sim} (\delta \tau)}}.
\label{sig}
\end{equation}
%

One can see the sequence of 
positive and negative peaks
with the significance
$\ga 4 \sigma$ with  respect to zero level.
Let us notice presence
of the long range ordering in dependence of
$\xi(\delta \tau)/\sigma(\xi)$ on $\delta \tau$.
We regard it as a consequence
of  the peak-and-dip structure
in the z-distribution of ALSs (see Sect.\ \ref{wavelet}).

The lower panel in Fig.\ \ref{XiSig}
represents the result of the power spectrum analysis
performed for 
the  value 
$\mu = \mu (\delta \tau) = \xi(\delta \tau)/\sigma(\xi)$
according to the equation: 
\begin{equation}
          {\rm P}_\mu (k)  =   {1 \over  N_\mu  D}
                          \left\{ \left[
          \sum_{m=1}^{N_\mu}  f_m  \cos \left(
          {2\pi k \, \delta \tau_m  \over \delta \hat{\tau}}
              \right) \right]^2  \, +  \, 
          \left[ \sum_{m=1}^{N_\mu}  f_m
             \sin \left(
       {2\pi k \, \delta \tau_m  \over  \delta \hat{\tau}}
              \right) \right]^2 \right\},
\label{Pk_mu}
\end{equation}
where 
$f_m=\mu_m - <\mu>, \,\, $
$\mu_m=\mu(\delta \tau_m), \,\, $
$<\mu>\ = N_\mu^{-1}\, \sum_m^{N_\mu} \mu_m \,\, $ 
is the mean value of $\mu$, 
$D =\ <\mu^2>-<\mu>^2 \, $ is the variance of $\mu$,
and $<\mu^2>\ = N_\mu^{-1}\, \sum_m^{N_\mu}\ \mu_m^2 \, $
is the mean squared value of $\mu$;    
the values $\delta \tau_m$   
run over a set of points 
$m=1,2,...N_\mu$  of the 
variable $\delta \tau$ from 0.1 to 1.86 
with the step 0.01. Thus
$N_\mu= 176$ is the full number of the points 
and  $\delta \hat{\tau} = 1.76$ 
is the whole interval of $\delta \tau_m$ 
variations under consideration.     
Note the strong single peak of P$_\mu (k)$ at $k \simeq 9$
on the significance level 
well exceeding  $5 \sigma$.
This peak manifests 
the periodicity 
of the correlation function
with the period
$\Delta \tau = 0.199 \pm 0.001$, which is in
a good agreement
with the periodicity
found in Sect.\ \ref{period}.

The one-dimensional correlation functions $\xi(\delta \tau)$ and
appropriate power spectra (\ref{Pk_mu}) were calculated also
for ALS samples with different values of 
$\delta v$ and $\Delta v$, 
where $\delta v$ is the averaging velocity interval
of ALS associations into single redshifts 
(see Sects.\ \ref{intro} and \ref{period}) 
and $\Delta v$
is the minimal velocity shift 
(along each line of sight)
adopted for the sample of ALSs  
relative to the 
original QSO
emission redshifts ($z_e$). 
In that way we exclude  
all ALSs  with $z_a$  within
the region associated with  their own QSOs, i.e.,
at $z_e \geq  z_a \geq  z_e - \Delta v \, (1+z_e)/c$.  
The values $\delta v$ and $\Delta v$  
were changed within the intervals
10 -- 10,000~km/s and 0 -- 10,000~km/s, respectively.
The results obtained are similar to that presented 
in Fig.\ \ref{XiSig}. Thus the results are 
quite robust to the variations of ALS samples.    

\begin{figure*}   
\vspace{-0.2cm}
\centering
\includegraphics[width=120mm,  bb=65  190  520  795,  clip]{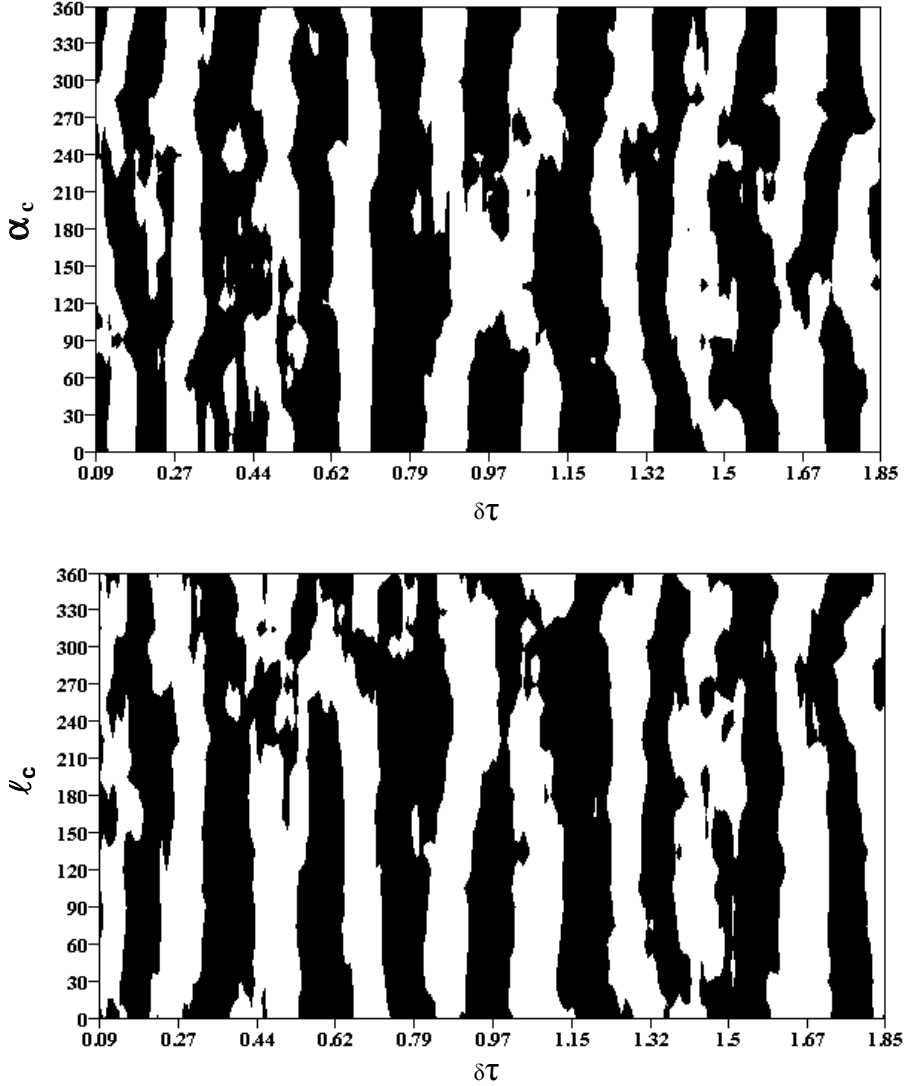}
\caption{
The correlation function $\xi(\delta \tau)$ 
of absorption systems
calculated
in 24 hemispheres 
for the equatorial ({\it upper panel}) 
and Galactic ({\it lower panel}) coordinate systems,
respectively;
$\xi > 0$ -- black strips and $\xi < 0$ -- white strips.  
{\it Upper panel}: 
the vertical axis indicates
the right ascension 
of the hemisphere centers $\alpha_c$ 
(in degrees).  
{\it Lower panel}:
the vertical axis indicates 
the Galactic longitudes of the hemisphere
centers $l_c$ (in degrees).
The horizontal axes show the interval
of $\delta \tau$        
between the components of ALS pairs. 
 }
\label{Xi-strips}
\end{figure*}
For comparison,  we have 
performed additional
calculations of the standard
two-point correlation function $\xi(r)$
(e.g., \citealt{ls93}, \citealt{jmst04}), 
where $r$ is the comoving distance
between the components of ALS pairs.  
The values 
$r$ were calculated using 
the formulae  by  \citet{r01} with 
equatorial coordinates $\alpha_i$, 
and  $\delta_i$  appropriate to ALSs
(Section\ \ref{isotr}).  
For convenience,
we employed 
the Hubble parameter $H(z)$
given by Eq.\ (\ref{H})
with $\Omega_{{\rm M}0}=0.3$  
and the same  coefficients
$A_0,\  A_1$ and $A_2$ as indicated in 
Section~\ref{period}. 
Let us  remind that 
the rescaling function (\ref{eta})
with the model dependence $H(z)$
displays 
the significant periodicity  
of the ALS redshifts (see Section~4).   
We tested a periodicity
of the function $\xi(r)$
(in a similar way as for $\xi(\delta \tau)$) 
using an analogue of Eq.\ (\ref{Pk_mu})
with the variable $\delta \eta = H_0 r/c $ 
instead of $\delta \tau$. 
As a result, we  have not revealed 
any significant peaks of P$_\mu (k)$
and, consequently, significant   
periodicities of the spatial 
correlation function $\xi(r)$.      

Fig.~\ref{Xi-strips}
shows two sets of the values
$\xi(\delta \tau)/\sigma(\xi)$ 
as a function
of  $\delta \tau$
calculated
for 24 hemispheres, 
i.e., the step of
consecutive rotations  
is also  $15^\circ$.
Similar to
Fig.~\ref{Strips}  
the upper and lower panels
of Fig.~\ref{Xi-strips} may be
treated as two-dimensional
distributions in the equatorial and Galactic coordinate 
systems, respectively.
The black strips display the positive  values of
the correlation function  ($\xi(\delta \tau)/\sigma(\xi)  >0$)
while the white strips display the negative  ones. 
One can see again (cf. Fig.~\ref{Strips})
that there is
the set of continuous vertical black and white
strips with centers at the same values $\delta \tau$
as on the upper panel of Fig.\ \ref{XiSig}.
It follows from comparison of the upper and lower panels
that the positions of the vertical strips as well as their
widths do not change essentially with transition 
from the equatorial to the Galactic
coordinates, i.e., 
with changing of the axis of hemisphere  
rotations.
Consequently, the two-point correlation
function (similar to the wavelet coefficients
in Sect. \ref{isotr}) 
is weakly sensitive to
orientations of the hemisphere.
  
\section{Distribution of Quasars and selection effects}
\label{QSOs}
The first peak 
at $z=1.95$ in the $z$-distribution of
emission and absorption line redshifts 
in QSO spectra
was found by \citet{bb67}. 
Then the quasi-periodical 
sequences of peaks and dips 
in the $z$-distribution
of quasar emission lines were  
marked by many authors
(e.g., \citealt{b68}, \citealt{c69}, \citealt{k71, k77, k90},
\citealt{fclc82},  \citealt{arpetal90}). 
Furthermore, several maxima and minima were detected 
in the distribution of absorption
systems (e.g., \citealt{cz89},  \citealt{arpetal90}).
Since the very beginning debates on availability
of the periodicities in the redshift distribution of QSOs
or QSO--galaxy pairs 
have been opened in the literature.
There are  many adherents
of the statement   
that some type 
of the periodicity 
does really exist
(e.g., \citealt{k90}, \citealt{kh90}, \citealt{arpetal90},
\citealt*{arpetal05},
\citealt{bn01}, \citealt{nb03}, \citealt{b03}, \citealt{be04}, 
and  references therein) and their opponents
(e.g., \citealt{br84}, \citealt{ba85, ba05}, \citealt{s91},
\citealt*{hmm02}, \citealt{tz05}),  
who argue that the periodicity and 
the very set of peaks and dips
in the QSO distribution 
are results of various selection effects.     

At least a part of the authors 
standing for reality 
of the periodicity  
under discussion
regards  that  as     
evidence for 
the hypothesis that QSOs
(or essential part of them) represent
some objects ejected from the nuclei of 
nearby active galaxies 
(e.g.,  \citealt{arpetal90, arpetal05}, \citealt{bn01},
\citealt{be04}, and references therein).
The redshifts of QSOs 
being  larger than the redshifts
of their parent galaxies have been 
assumed to have a non-cosmological
``intrinsic'' origin   
and display a set of 
preferred
(discrete) quasi-periodical values.
To our knowledge,
there are  
two models 
discussed in literature
which suggest different  
periodical sets
of preferred redshifts. 
The first model has been applied mainly
to QSOs  and marginally to ALSs
in QSO spectra  (see above).  
That  is described by
so called Karlsson
formula for the redshift periodicity
$\Delta \log (1+z_e)=0.089$ 
(e.g., \citealt{k71, k77, k90}, 
\citealt{arpetal90, arpetal05}, 
\citealt{bn01}, \citealt{nb03}). 
The second one  was proposed
by Bell (e.g.,  \citealt{be02, be04}) 
for  the ``intrinsic'' redshifts 
of QSOs 
and extended on a set of
preferred redshifts of galaxies
by \citet{beco03}.
According to both  
scenarios, the formation of ALSs    
on the lines of sight to the QSOs  
should have also occurred in the local vicinity
of the Universe. 

To test a possible correlation 
of the ALS distribution
with expectations of the first model
we have calculated the
power spectrum of the sample of ALSs
with respect to
the trial function $\log (1+z)$.
We have found no significant power peaks 
which would have a chance to be related with
the period 0.089 or with a multiple value of it.
According to the second QSO ejection model,    
appearance of the preferable $z_{\rm max}$
and $z_{\rm min}$ in the ALS distribution   
could be, in principle, interpreted 
as effects of an additional set 
of discrete  
``intrinsic'' redshifts 
referred  to galaxies.
To test this statement one can compare
the values of $z_{\rm max}$  from
the last but one column in Table~1
with 
the set of intrinsic redshift components 
$z_{iG}[N,\ m]$ 
(where $N$ and $m$ are some  quantum numbers) 
defined for galaxies 
by  Eq.\ (B1)  of  \citet{beco03}. 
Such a comparison shows that 
the set of the peaks $z_{\rm max}$
given in Table~1 are not consistent 
with the intrinsic redshifts
of galaxies.  
Only three values $z_{\rm max}$
turn out to be equal
(within the uncertainty $\pm 0.02$)
to the values  $z_{iG}$ 
from the set of 30 numbers at $N \leq 8$
within the interval $0.1 \leq z  \leq 3.7$.  
Thus we have found no traces of consistency
between our results and the hypotheses 
of non-cosmological 
``intrinsic'' redshifts of QSOs.
Accordingly,  
in contrast with the papers quoted above ,  
we assume that the redshift of 
the absorption lines $z_a$ is cosmological
and the observed ALSs are associated 
with ionized gas in intervening 
galaxies or clusters of galaxies
at cosmological distances along lines of sight
to original QSOs. 

The results of this paper 
may be  biased 
by  selection effects 
appropriate to the sample of ALSs.  
Taking it into account 
we undertook 
some special efforts 
in Paper~I
to minimize 
such effects in the statistical analysis.  
As it is shown in Table~1 the results 
obtained in Paper~I  are 
consistent with the results of the present paper. 
In Section~\ref{CF} 
we discuss
the effects of excluded regions of $z$  
(within $\Delta v$) 
near QSO emission redshifts. 
Such exclusions allow us to avoid
at least a part of
selection effects related to 
so called associated regions of QSOs
(e.g., \citealt{qby96}).    

Furthermore, 
we  failed to find 
similar periodicity 
(at significance level $\ga 3 \sigma$)
in the $z$-distribution 
of the  661  original QSOs 
as it is revealed in the  distribution of ALSs    
registered in their spectra
(see Section~\ref{period}). 
It separates possible selection effects
biasing the statistics of QSOs 
(e.g., \citealt{br84}, \citealt{ba05}, \citealt{tz05})
and those peculiar to
the sample of ALSs involved in the analysis.  
It is worthwhile to note that 
the characteristic scale of redshifts $z$ 
which we consider
in this paper ($\sim 0.18$) is essentially smaller 
than the scale ($\sim 0.48$) discussed
for  QSOs  and  QSO--galaxy pairs  
(e.g., \citealt{bn01}) 
or the $z$-period ($\sim 0.62$) 
found  for a sample of QSOs  
by \citet{be04}.
In particular, 
the narrowness 
of the peaks and dips 
obtained here
makes  the analysis of \citet{ba79, ba83}
less relevant to our results.

\begin{figure*}    
\centering
\includegraphics[width=120mm,  bb=60  465  520  780, clip]{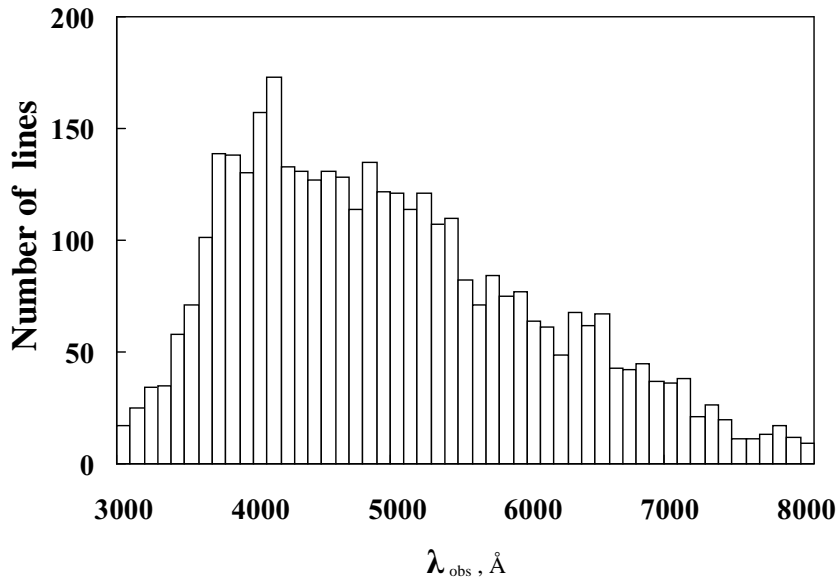}
\vspace{-0.2cm}
\caption{
$\lambda$-distribution
(spectroscopic completeness) 
of resonance doublets and single absorption lines
of 15 ions which contribute mostly
in the distribution of ALSs 
at different $z$;  bin size of
the histogram 
is $\Delta \lambda = 100 \AA$.    
}
\label{Lines}   
\end{figure*}
In Fig.~\ref{Lines} we 
test at least a part of possible selection
effects concerned with  
{\it a priori} different
probability 
to register absorption lines  
in  QSOs spectra  
at different values of wavelength $\lambda$ 
within the observational interval (3000 -- 8000 $\AA$).
We choose 15 single lines or resonance doublets  
of the ions 
(C~II,\ C~IV,\ Mg~I,\ Mg~II,\ Si~II,\ Si~III,\ Si~IV,\ 
N~V,\ Al~II, Al~III,\ Fe~II)
most representative in the sample of ALSs
under consideration and   
determine a number of wavelengths   
$\lambda=\lambda_0 (1+z_a)$
observed
for each of them, where     
$\lambda_0$ is a laboratory value. 
Fig.~\ref{Lines} represents
a histogram   
of the absorption lines 
as a result of count of the lines
within bins  $\Delta \lambda = 100 \AA$.  
One can see quite smooth histogram 
(so-called spectroscopic completeness)
with two  appreciable 
peaks between $4000$ -- $4200\ \AA$
and $6300$ -- $6600\ \AA$
(at significance level $\ga 2 \sigma$).
These peaks are obviously not enough 
to explain all set of peaks  
in Fig.~\ref{AS91-03}
as pure selection effects, i.e., as a consequence 
of peaks 
in the spectroscopic completeness.
Although this conclusion does not completely exclude 
other selection  effects in our statistical 
consideration.
         
\section{Conclusions and discussion}
\label{concl}
The main results of the 
statistical analysis  of 2003
absorption-line systems (ALSs)  
in the redshift range $z = 0.0-3.7$
can be summarized as follows:  

(1)   The z-distribution of ALSs displays
the statistically significant pattern 
of alternating maxima (peaks) and minima
(dips) relative to a smooth  curve.   
The positions of the maxima and minima 
(centroids of the 
peaks and dips) in the $z$-distribution are
given in Table~1 with an accuracy  $\pm 0.02$.
These positions are predominantly not sensitive  
(within statistical uncertainties)
to orientations of the hemispheres under 
statistical consideration.        
However, the correlation coefficients
between the $z$-distributions 
calculated for 12 pairs of opposite 
(independent) hemispheres
turn out to be moderately significant
($\ga 2 - 3 \sigma$) as a consequence of 
variations of the  peak-and-dip  profiles 
at relatively stable weighted centers  
of the peaks and dips. 
 
(2)   The sequence of peaks and dips 
reveals a certain regularity.
The power spectrum  calculated
according to Eq.\ (\ref{P_k})
with the rescaling function $\tau(z)$
given by  Eqs.\ (\ref{tau}) and (\ref{eta})   
displays the peak for the harmonic number $k=16$ 
(period  $\Delta \tau = 0.199 \pm 0.001$)
at the significance level exceeding $4 \sigma$ 
(relatively to the hypothesis 
of the uniform distribution of ALSs over $\tau(z)$).
The rescaling function $\tau(z)$ 
may be treated as a measure 
of temporal separations of the epochs
with different $z$.
Still more prominent peak 
at the same period $\Delta \tau$ arises
in the power spectrum  calculated
for the two-point correlation function
$\xi (\delta \tau)$  with the use of Eqs.\ (\ref{xi})  
and  (\ref{Pk_mu}). 
   
(3)  The obtained distribution of ALSs is likely to be
coupled with the appearance of alternating pronounced (peaks)
and depressed (dips) epochs in the course of the cosmological
evolution, i.e., with the existence of some (relatively weak)
spatial-temporal wave process. 
According to
the cosmological principle (e.g, \citealt{p93})
similar wave-like process 
would be observed
from any spatial-temporal points in the Universe.

\begin{figure*}    
\vspace{-0.2cm}
\includegraphics[width=120mm,  bb=60  440  580  790,  clip]{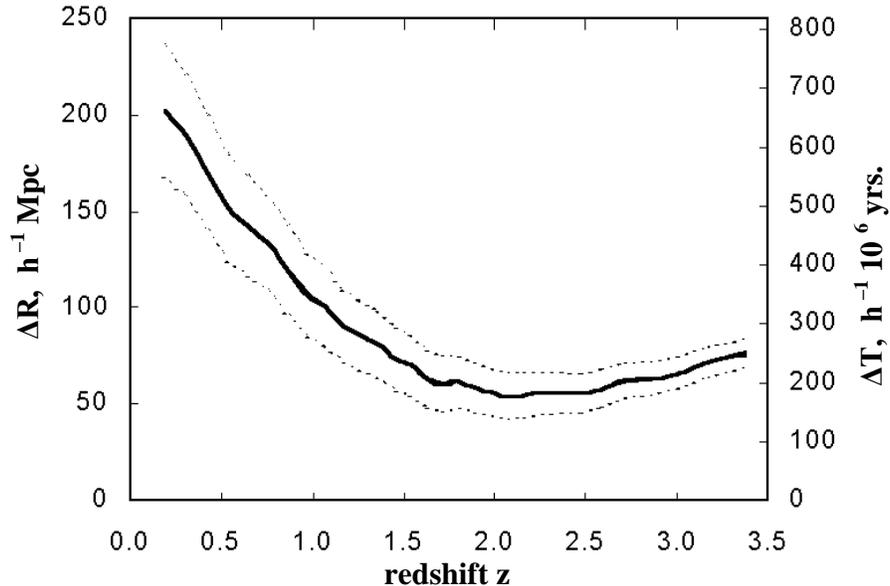}
\caption{
Comoving distance intervals $\Delta R$ (left vertical axis)  
and appropriate time intervals $\Delta T$ (right vertical axis)   
between successive neighbour peaks in the $z$-distribution of ALSs
calculated  as  
the intervals  $\Delta \eta$   
multiplied by $c/H_0$ and $1/H_0$, respectively 
(see text),  
and referred to  appropriate
redshifts $z$.
The dotted lines illustrate $\pm \sigma$ 
deviations.
}
\label{RT-scales}
\end{figure*} 
(4)    Fig.~\ref{RT-scales} demonstrates   
the intervals of
the comoving distance, 
$\Delta R = c \Delta \eta/H_0$, and
the appropriate time,  
$\Delta T = \Delta \eta/H_0$, 
between  neighbour peaks in dependence 
on the current redshifts $z$, 
where 
$\eta(z)$ is given by Eq.\ (\ref{eta}),
$\Delta \eta$ runs over 
all  successive intervals 
$\eta(z_{\rm max}^{l+1}) - \eta(z_{\rm max}^l)$,
$\,  l=1, 2, ...$ -- numerates the  peaks in the 
$z$-distribution of the ALSs 
(from top to bottom in Table~1)\footnote{Note that the
values $\Delta R$ and $\Delta T$ were calculated
with normalization of $\int_0^\infty {\rm d} z$
in Eq.~(\ref{eta}) to unity}.
It follows from Fig.~\ref{RT-scales} 
that the characteristic scales 
of  $\Delta R$ 
and   $\Delta T$
vary in the ranges
$\Delta R = (50$ -- $200)\, h^{-1} $~Mpc  and
$\Delta T = (150$ -- $650)\, h^{-1}$~Myr,
respectively, where $h=H_0/100$~km(s~Mpc)$^{-1}$.

Let us note that 
our treatment of the results obtained  
does not contradict to the existence of the  
Large Scale Structure (LSS) of the
matter distribution in the Universe (e.g.,
\citealt{sz89}, 
\citealt{ein02}, \citealt{tseema02}, 
\citealt{dd04}, \citealt{jmst04}, \citealt{eisen05}  
and references therein).   
The spatial-temporal variations discussed here 
may be superimposed 
upon the process of formation and evolution
of the LSS elements. 
It was noted in Sect.\ \ref{isotr} 
that the strip-like picture represented 
in Fig.~\ref{Strips}  
becomes more fragmentary  if
the $z$-distributions are calculated for  the
quadrants ($\alpha_c \pm 45^\circ$)
instead of the hemispheres.   
One can assume that more  
fragmentary patterns  
obtained for smaller sectors of observations  
appear not only as a result 
of a reduction of ALS  statistics
but also due to  
an interplay between  LSS 
and the variations  
discussed here. In any case
the possibility of such an interplay 
deserves a special investigation.  

Extended discussions 
of the effects of the 
periodicity in the redshift distribution 
of galaxies on a scale about $130~h^{-1}$~Mpc  
was initiated  in literature
by the results of a pencil-beam survey 
of \citet{beks90}.    
These results also 
admit both interpretations
either as
a pure spatial quasi-periodic 
pattern of LSS  constituents 
(set of clumps or walls and voids;
e.g.,  \citealt{kp91}, \citealt{vdw91},  \citealt{dbps92}, 
\citealt{yosh01} and references therein)
or as a temporal 
sequence of  
pronounced and depressed epochs
becoming apparent
in the matter distribution.
The latter treatment is closer
to our interpretation of 
the results represented here.
However, our statistical approach 
does principally
differ from the pencil-beam  consideration.

The appearance of alternate 
pronounced and depressed epochs
may be associated with intrinsic
quasi-periodical properties of galaxies 
and/or their clusters
(e.g., \citealt{lh98}).
On the other hand, 
it  can be explained
as an apparent pattern of density variations
in terms of the cosmological
scenarios considering
scalar (scalar-tensor) 
fields and/or phase transitions
at different evolutionary stages.
Such effects may lead to 
temporal oscillations
of the Hubble parameter $H(z)$  
(e.g., \citealt{m90,  m91},
\citealt*{hst90}, \citealt{ps02}) and/or
effective gravitational constant $G$ 
(e.g., \citealt{sv94}, \citealt{bcetal94}, 
\citealt{gqss01},
\citealt*{bps03}, \citealt{d05}),
peculiar velocity field \citep{hst91},
cosmological constant $\Lambda$   
(e.g., \citealt{dp94}, \citealt*{dks00}),
and, as a consequence, to apparent oscillations 
of the matter distribution with the redshift.

The rapid development of ground-based and space-born
observational facilities in the optical, infrared,
and near-ultraviolet ranges in recent years gives confidence
that the hypothesis of the existence of the alternate epochs
with enhanced and reduced concentrations of 
absorbing (or luminous) matter 
in the Universe may be tested in the nearest future.

\textit{Acknowledgments}
We are grateful to V.S. Beskin for suggestion
to apply the wavelet analysis. We are also grateful 
both anonymous referees for useful critical remarks.
The work has been supported partly
by the RFBR (grant No.\  05-02-17065a), 
and  by the Federal Agency for Science and
Innovations  (grant NSh\ 9879.2006.2).

\end{document}